\documentclass[envcountreset,oribibl,orivec]{llncs}
\usepackage{graphics}
\usepackage{epsfig}
\usepackage{pifont}
\usepackage{amsfonts}
\usepackage{amssymb}
\usepackage{amsmath}

\makeatletter
  \@addtoreset{equation}{section}
  \renewcommand{\theequation}{\thesection.\@arabic\c@equation}
  \makeatother

\begin{document}

\title{Modelling the\\ Biomacromolecular Structure\\with Selected
Combinatorial\\Optimization Techniques}
\author{R. P. Mondaini}
\institute{Federal University of Rio de Janeiro\\Alberto Luiz
Coimbra Institute for Graduate Studies
\\and Research in Engineering \\COPPE/UFRJ - Technology Centre \\21.941-972 - P.O. Box 68511,\\ Rio de Janeiro,
RJ, Brazil\\\email{rpmondaini@gmail.com, mondaini@cos.ufrj.br}}
\maketitle

\begin{abstract}
Modern approaches to the search of Relative and Global minima of
potential energy function of Biomacromolecular structures include
techniques of combinatorial optimization like the study of Steiner
Points and Steiner Trees. These methods have been successfully
applied to the problem of modelling the configurations of the
average atomic positions when they are disposed in the usual
sequence of evenly spaced points along right circular helices. In
the present contribution, we intend to show how these methods can
be adapted for explaining the advantages of introducing the
concept of a Steiner Ratio Function (SRF). We also show how this
new concept is adequate for fitting the results obtained by
computing experiments and for providing an improvement to these
results if we use the restriction of working with Full Steiner
Trees.
\end{abstract}

\section{Introduction}
The study of Steiner Trees was shown to be useful for
understanding the structure of biomacromolecules
\cite{sm:tree,mo:bio1,mo:bio2,mo:bio3,mo:bio4}, through the
application of combinatorial optimization techniques. These
methods have been usually applied in the modelling of
configurations of evenly spaced points along right circular
helices since these are adequate to fit the average atomic
positions. Some writers have failed in their efforts at
correlating the Steiner Ratio with the potential energy of a given
special configuration. They didn't succeed in this trial to
develop a robust method of Global Optimization. In the present
work we propose to extend the concept of a Steiner Ratio into that
of a SRF \cite{mo:kluw}. This was possible by working with
experimental data obtained from a modification of the W. D.
Smith's algorithm \cite{si:algo}, which was adapted to work with
only one topology. In our experiments, we have followed the
prescription of using the topology of the sausage configuration
\cite{sm:ratio}. The study of the candidates for SRF is part of a
full geometric approach to the problem of macromolecular
structure. It has deep connections with the problem of a good
definition of chirality measure \cite{mo:gmb}, as well as it
provides nice insights in the understanding of the homochirality
phenomena and its importance for the stability of macromolecular
configurations. It seems that chirality of biomacromolecules has
so a fundamental importance that its study deserves the use of
more powerful methods than those usually considered by
Combinatorial Optimization. In our first approach, we proposed a
chirality function as a constraint in a thermodynamically inspired
idea of constructing a new Cost Function as a Gibbs Free Energy
\cite{mo:gmb,mo:bio3}. In a more elaborate theory, the aspects of
potential energy of the configuration and its chirality should
come from a self-contained Steiner Function.

There will be no need for a constraint. At the present stage of
our research, we have to pave the way for this future theory by
undertaking the study of possible candidates for SRF. We have also
taken into consideration the restrictions imposed by the natural
requirement of full Steiner Trees.

\section{The Steiner Ratio}
Let us consider a finite set of points $A$ in a metric manifold
$M$. We consider all the possible ways ($s$-topologies) of
connecting pairs of points on each set of this manifold. The
resulting edges are supposed to be geodesics of the manifold and
their collection is a tree. We get a spanning tree (SP) by
discarding the edge of greatest length. Among these spanning trees
of the set $A$ with different $s$-topologies and length
$l_{\mathrm{SP}}(s,A)$, there is one which overall length is a
minimum as compared to all the trees of the same set. This is the
minimum spanning tree of the set $A$, MST(A), and its length is
\begin{equation}
l_{\mathrm{MST}}(A)=\min_{(s-\,\mathrm{topologies})}\,l_{\mathrm{SP}}(s,A)\enspace
.
\end{equation}

If we now allow for the introduction of additional points on each
set $A$ of the manifold in order to have spanning trees of smaller
overall length, we shall have the concept of a Steiner tree (ST).
In the construction of these trees we have to follow the
additional requirement of the tangent lines to the geodesic edges
meeting at $120\,^{\circ}$ on each Steiner (additional) point
($t$-topologies). Among these Steiner trees of the set $A$ with
different $t$-topologies and length $l_{ST}(t,A)$, there is one
which overall length is a minimum. We call it the Steiner Minimal
Tree of the set $A$, $SMT(A)$. Its length is given by
\begin{equation}
l_{\mathrm{SMT}}(A)=\min_{(t-\,\mathrm{topologies})}\,l_{\mathrm{ST}}(t,A)\enspace
.
\end{equation}

The $\mathrm{MST}(A)$ is considered as the worst approximation
(the ``worst cut") to the $\mathrm{SMT}(A)$. It is usual to
associate a number to this couple of minimal trees of the set $A$:
the ratio of their overall lengths. This is calculated with the
definition of distance of the manifold $M$. This number is called
the Steiner Ratio of the set $A\subset M$. We write,
\begin{equation}
\rho(A)=\frac{l_{\mathrm{SMT}}(A)}{l_{\mathrm{MST}}(A)}\enspace .
\end{equation}

The Steiner Ratio of the manifold $\rho_M$ is then defined to be
the infimum of the sequence of values $\rho(A)$, or
\begin{equation}
\rho_M=\inf_{A\,\subset\, M}\,\rho(A)\enspace .
\end{equation}

\section{The Steiner Ratio Function}

The concept of a Steiner Ratio Function is a specialization of the
definitions above. It is better introduced by examples. We now
suppose that all the sets $A$ of the manifold $M$ have the same
number $n$ of points. Moreover, we also take these points $P_i$ to
be evenly spaced points along a right circular helix. The
cartesian coordinates of the sequence of consecutive points is
given by

\begin{equation}
P_i(\cos{i\omega},\sin{i\omega},\alpha
i\omega),\,\,\,\,\,\,\,0\leq i\leq n-1
\end{equation}
where $2\pi \alpha$ is the pitch of the helix.

Equation (3.1) means that we have a helical point set for each
pair of values $(\omega,\alpha)$. We now consider the subsequences
of points obtained from (3.1) by skipping points. They are of the
form $P_{j+kl}$, with $j=0,1,2,\ldots,k-1$; $(k-1)=$ number of
skipped points; $l=$ number of intervals of skipped points before
the present point. There are $n$ possible sequences. Among these
only $k$ of them contain different points. The process of
construction of these subsequences is the following:

Given a number of $n$ points evenly spaced along a right circular
helix, we form a $j-$subsequence with $(k-1)$ skipped points, and
a maximum $l-$value. Since the index $(j+lk)$ should be restricted
by
\begin{equation}
j+lk\leq n-1\enspace .
\end{equation}

We have
\begin{equation}
l_{\mathrm{max}}=\left[\frac{n-j-1}{k}\right]\enspace .
\end{equation}

Where $[\,.\,]$ stands for the greatest integer value. A
$j-$subsequence is then given by
\begin{equation}
\left(P_j\right)_{k\,l_{\mathrm{max}}}:P_j,P_{j+k},P_{j+2k},\ldots,P_{j+l_{\mathrm{max}}k}\enspace
.
\end{equation}

We also define a sequence which is the union of all the
$j-$subsequences by defining connection edges of the highest end
and/or lowest end consecutive points
\begin{equation}
\bigcup_{j=0}^{k-1}\left(P_j\right)_{k\,l_{\mathrm{max}}}\enspace
.
\end{equation}

Some examples will be worth to illustrate the results above.

{\bfseries 1.} Let $n=23$, $k=2$, $j=0,1$.

We shall have
\[l_{\mathrm{max}}(j=0)=11, l_{\mathrm{max}}(j=1)=10\]
and
\[P_{j+l_{\mathrm{max}}k}=P_{22},\,P_{21}\,,\,\,\,\,\mbox{respectively}.\]

The subsequences are
\begin{eqnarray}
\left(P_0\right)_{2,11}&:&P_0,P_2,P_4,P_6,P_8,P_{10},P_{12},P_{14},P_{16},P_{18},P_{20},P_{22}\\
\left(P_1\right)_{2,10}&:&P_1,P_3,P_5,P_7,P_9,P_{11},P_{13},P_{15},P_{17},P_{19},P_{21}\enspace
.
\end{eqnarray}

{\bfseries 2.} Let $n=23$, $k=3$, $j=0,1,2$.

We have
\[l_{\mathrm{max}}(j=0)=7, l_{\mathrm{max}}(j=1)=7, l_{\mathrm{max}}(j=2)=6\]
and
\[P_{j+l_{\mathrm{max}}k}=P_{21},\,P_{22},\,P_{20}\enspace .\]

The subsequences are now
\begin{eqnarray}
\left(P_0\right)_{3,7}&:&P_0,P_3,P_6,P_9,P_{12},P_{15},P_{18},P_{21}\\
\left(P_1\right)_{3,7}&:&P_1,P_4,P_7,P_{10},P_{13},P_{16},P_{19},P_{22}\\
\left(P_2\right)_{3,6}&:&P_2,P_5,P_8,P_{11},P_{14},P_{17},P_{20}\enspace
.
\end{eqnarray}

The sequence corresponding to (3.1) can be recovered by making
$j=0$, $k=1$ in the process described above, it can be also
written as $\left(P_0\right)_{1,(n-1)}$.

To each sequence of the form (3.5) corresponds a spanning tree
$\mathrm{SP}(k,\omega,\alpha)$. In Fig.1, we show the sequence of
points of (3.1) and of the two examples above, respectively.
\pagebreak
\begin{figure} [hp]
\hfil\scalebox{0.8}{\includegraphics*{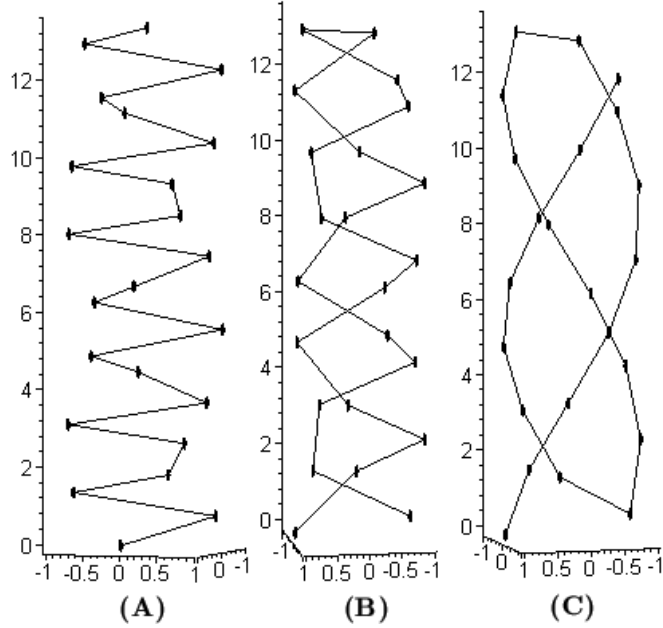}}\hfil
\caption{{\bfseries (A)} The sequence $n=23$, $k=1$, $j=0$;
{\bfseries (B)} The union of the sequences $n=23$, $k=2$, $j=0$
and $n=23$, $k=2$, $j=1$; {\bfseries (C)} The union of the
sequences $n=23$, $k=3$, $j=0$; $n=23$, $k=3$, $j=1$ and $n=23$,
$k=3$, $j=2$.} \label{f1}
\end{figure}

In the following, we take $M$ to be the $\bbbr^3$ manifold with an
Euclidean definition of distance, or $E^3$.

The coordinates of the points $P_{j+kl}$ are given analogously to
(3.1). The Euclidean length of $n-$point configurations like those
of Fig.1 above can be written,
\begin{equation}
l_{\mathrm{SP}}(k,\omega,\alpha)=(n-k)\sqrt{k^2\alpha^2\omega^2+A_k+1}+(k-1)\sqrt{\alpha^2\omega^2+A_1+1}
\end{equation}
where
\begin{equation}
A_k=1-2\cos(k\omega)\enspace .
\end{equation}

There is only one Steiner Tree for all these point configurations.
It has the 3-sausage's topology \cite{si:tre2} and its Euclidean
length \cite{mo:kluw,mo:bio2,mo:tema} is written as
\begin{equation}
l_{ST}(\omega,\alpha)=(n-2)(1-r)+(n-3)\alpha\omega\sqrt{\frac{A_1+1}{A_1}}+2\sqrt{\alpha^2\omega^2+(1-r)^2+r(A_1+1)}
\end{equation}
where
\begin{equation}
r(\omega,\alpha)=\frac{\alpha\omega}{\sqrt{A_1(A_1+1)}}\enspace .
\end{equation}

The Steiner points are also in a helix of the same pitch but
smaller radius $r(\omega,\alpha)$ as compared to the points of the
configuration (3.1).

In order to adapt the formulae (2.1) and (2.2) to the present case
of a SRF definition we should note that if we take $n\gg k$, all
the sets $A\subset E^3$ should be considered as the same. This is
a set of a great number of points evenly spaced in a right
circular helix.

For $n\gg k$, we then have, from (2.2), (3.13) and (3.14),
\begin{equation}
l_{\mathrm{SMT}}(\omega,\alpha)=n\left(1+\alpha\omega\sqrt{\frac{A_1}{A_1+1}}\right)
\end{equation}
and from (2.1), (3.11) and (3.12),
\begin{equation}
l_{\mathrm{MST}}(k,\omega,\alpha)=n\sqrt{k^2\alpha^2\omega^2+A_k+1}\enspace
.
\end{equation}

The Steiner Ratio Function is now defined, according (2.3), (3.15)
and (3.16) as
\begin{equation}
\rho(\omega,\alpha)=\frac{1+\alpha\omega\sqrt{\frac{A_1}{A_1+1}}}{\underset{k}{\min}\left(\sqrt{k^2\alpha^2\omega^2+A_k+1}\right)}\enspace
.
\end{equation}

The ``$\underset{k}{\min}$" in the equation above should be
understood in the sense of a new function formed in a piecewise
way from the functions corresponding to the chosen $k-$values. For
$k=1,2,3$ this means
\[\min\left(\sqrt{\alpha^2\omega^2+A_1+1},\sqrt{4\alpha^2\omega^2+A_2+1},\sqrt{9\alpha^2\omega^2+A_3+1}\right)\enspace.\]

The associated Steiner Ratio for these helical point
configurations will be given by the Global minimum of the function
$\rho(\omega,\alpha)$ above, according (2.4).

\section{The Fitting of Computational Results of the W. D. Smith's Algorithm}

In Fig.2 below, we represent a section of the SRF given by (3.17)
for $\alpha=\alpha_\mathrm{R}=0.26454000216\ldots$. This is the
pitch of a helical configuration in which the points are regular
tetrahedra vertices. These tetrahedra are glued together at common
faces to form a 3-sausage configuration \cite{du:disp}. The
condition for equal edges of the tetrahedra lead to the following
equations:

\begin{equation}
\alpha\omega=\frac{\sqrt{3}}{3}\sqrt{A_1-A_2},
\end{equation}

\begin{equation}
\alpha\omega=\frac{\sqrt{5}}{5}\sqrt{A_2-A_3},
\end{equation}

\begin{equation}
\alpha\omega=\frac{\sqrt{2}}{4}\sqrt{A_1-A_3}.
\end{equation}

Only two of them are independent and the non-trivial solution is

\begin{equation}
\omega=\pi-\arccos\left(\frac{2}{3}\right)=
2.30052398302\ldots,\,\,\,\alpha=\frac{\sqrt{30}}{9(\pi-\arccos\left(\frac{2}{3}\right))}=0.26454000216\ldots
\end{equation}

We take $k=1,2,3$ and the points obtained from the W. D. Smith's
algorithm \cite{si:algo} with a search space reduction, i.e.,
adopting the 3-sausage's topology as the only feasible. The
modified algorithm is available at the site www.biomat.org.

\begin{figure} [hp]
\begin{center}
\scalebox{0.8}{\includegraphics*{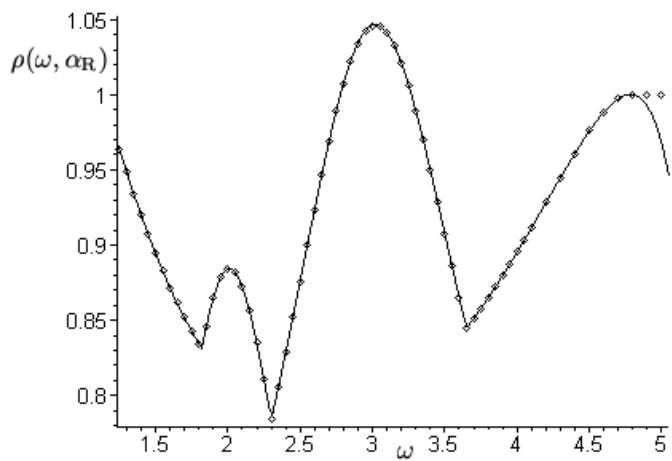}}
\end{center}
\caption{The section $\alpha=\alpha_{\mathrm{R}}$ of the surface
$\rho(\omega,\alpha)$ (---\,) for $k=1,2,3$ with points obtained
from computing experiments ($\diamond$).} \label{f4}
\end{figure}

It should be noted from Fig.2 that the function (3.17) is a very
good fitting to the experiments which were done with the W. D.
Smith's algorithm. It is so a good fitting, that it has the same
bad performance at some $\omega-$region $(\rho>1!)$ as can be seem
in Fig.2 above. Although these can be easily explained by the
difficulty of working with a quasi-plane configuration (the
neighbourhood of $\omega=\pi$) when we choose a priori
3-dimensional configurations, we prefer to circumvent this
difficulty in the next section. We do it by introducing some
necessary restrictions related to Full Steiner Trees.

From the viewpoint of the search of a Global minimum, the surface
given by (3.17) is also a very good candidate as can be seen from
its level curves in Fig.3 below \pagebreak

\begin{figure} [hp]
\hfil\scalebox{0.8}{\includegraphics*{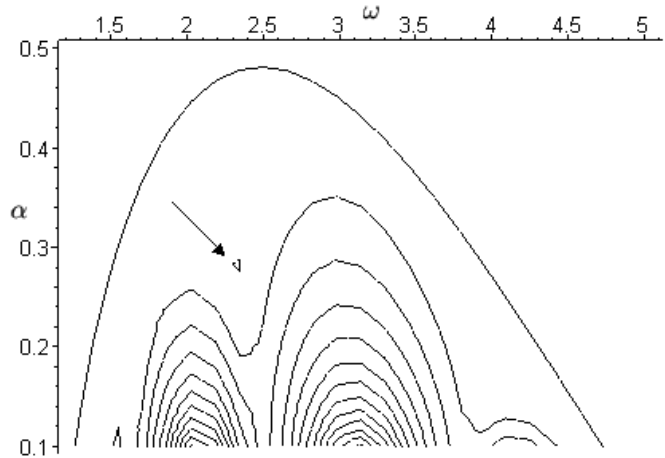}}\hfil
\caption{Level curves of the surface $\rho(\omega,\alpha)$,
(3.17). The region which contains the projection of the Global
minimum point is pointed out by the arrow.} \label{f5}
\end{figure}

If we substitute into (3.17) the values $(\omega,\alpha)$
corresponding to the 3-sausage's configuration, or
$\omega=\pi-\arccos(2/3)$, $\alpha=\sqrt{30}/9(\pi-\arccos(2/3))$,
we get

\begin{equation}
\rho=\frac{1}{10}(3\sqrt{3}+\sqrt{7})=0.78419037337\ldots
\end{equation}

which is the value given by the authors of \cite{du:disp} for the
best upper value for the Steiner Ratio of the $E^3$-manifold.

\section{The Restriction to Full Steiner Trees}

In the last section there is no restriction to Full Steiner Trees.
We take as an assumption that the natural organization of
biomacromolecular structure is biased by Full Steiner Trees. A
structure is non-degenerate or it is almost completely degenerate.
Nature can provide a process in which the Steiner Ratio is
approaching continuously to the value $\rho=1$. However, the tree
cannot be partially degenerate in the sense of partial full trees
connected at degenerate vertices.

In this section, we emphasize that we need more stringent
constraints, instead of a constraint relaxation for a good
definition of SRF. ``Good" means here the definition which leads
to $\rho-$values lesser than the value reported in Section 4 if we
believe that the main conjecture of the authors of \cite{sm:ratio}
could be disproved.

If we look at Fig.2 above, we can see that the algorithm used as
well as the modelling based on (3.17) do not make any
discrimination to degenerate Steiner Trees. This is due to the
fact that there are regions of $\omega$-values in which $\rho$ is
close to 1. To each $(\omega,\alpha)$ pair of values there is
associated a point configuration. This means that there are
regions in the line $(\omega,\alpha_{\mathrm{R}})$ corresponding
to degenerate configurations. This is also valid for other
$\alpha-$values since these sections of the surface (3.17) have
similar profiles.

We can now introduce the restriction to full Steiner Trees. Let us
take the $j$, $k$ subsequence of (3.4). The points
$P_j,P_{j+k},P_{j+2k},\ldots$ are evenly spaced along the
corresponding right circular helix. The angle made by contiguous
edges is
\begin{equation}
\cos{\theta_k}=\frac{\overrightarrow{P_{j+lk}P_{j+(l+1)k}}\cdot\overrightarrow{P_{j+(l+2)k}P_{j+(l+1)k}}}{\|\overrightarrow{P_{j+lk}P_{j+(l+1)k}}\|\|\overrightarrow{P_{j+(l+2)k}P_{j+(l+1)k}}\|}
\end{equation}
where $\|\cdot\|$ is the Euclidean norm.

The cartesian coordinates of the points $P_{j+kl}$ can be written
analogously to (3.1) and we have
\begin{equation}
\cos{\theta_k}=-1+\frac{(A_k+1)^2}{2(k^2\alpha^2\omega^2+A_k+1)}
\end{equation}
where $A_k$ is given by (3.12).

The restrictions to Full Steiner Trees can be then written in the
form
\begin{equation}
\cos{\theta_k}\geq -\frac{1}{2}\enspace .
\end{equation}

In Fig.4 below we show the restrictions introduced by (5.3) for
$k=1,2,3$. The horizontal line corresponds to the value
$\cos{\theta_k}=-1/2$.

\begin{figure} [hp]
\hfil\scalebox{0.8}{\includegraphics*{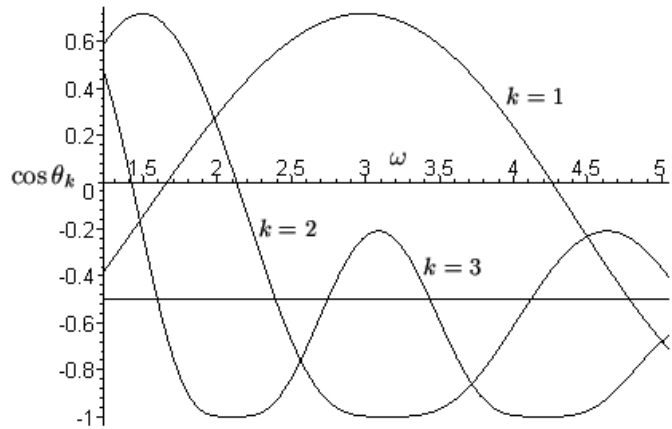}}\hfil
\caption{The restriction to Full Steiner Trees. The figure is a
section of the surfaces (5.2) for $\alpha=\alpha_{\mathrm{R}}$.}
\label{f6}
\end{figure}

The spanning trees corresponding to $k=2$ have a large forbidden
region. The $k=1$ curve is the only one which corresponds to Full
Steiner Trees in a large region of the $\omega-$interval studied
in this work. The proposal for the Steiner Ratio Function is given
in this case by
\begin{equation}
\rho_1(\omega,\alpha)=\frac{l_{\mathrm{SMT}}(\omega,\alpha)}{l_{\mathrm{MST}}(1,\omega,\alpha)}\enspace
.
\end{equation}

From (3.17), we get
\begin{equation}
\rho_1(\omega,\alpha)=\frac{1+\alpha\omega\sqrt{\frac{A_1}{A_1+1}}}{\sqrt{\alpha^2\omega^2+A_1+1}}\enspace
.
\end{equation}

Fig.5 shows the sections $\alpha=\alpha_{\mathrm{R}}$ of the
surfaces given by (3.17) and (5.5). By restricting our
observations to the $\omega-$interval in which the $k=1$ curve in
Fig.4 allows for Full Steiner Trees, it should be noted that the
section $\alpha=\alpha_{\mathrm{R}}$ of the candidate for a SRF
given by (5.5) is the convex envelope \cite{mk:prog} of the
section $\rho(\omega,\alpha_{\mathrm{R}})$ of (3.17).

\begin{figure} [hp]
\hfil\scalebox{0.8}{\includegraphics*{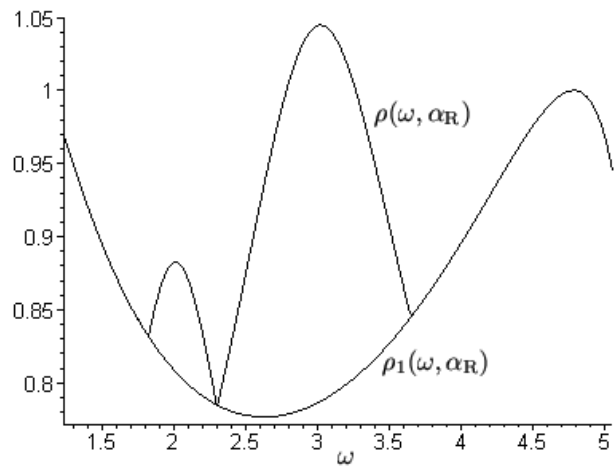}}\hfil
\caption{The $\rho_1(\omega,\alpha_{\mathrm{R}})$ candidate is a
convex envelope of the function
$\rho(\omega,\alpha_{\mathrm{R}})$.} \label{f7}
\end{figure}

\section{Concluding Remarks}

The restriction of the point configurations only to those which
lead to Full Steiner Trees, has wiped out the Global minima of the
surface $\rho(\omega,\alpha)$. The resulting surface
$\rho_1(\omega,\alpha)$ satisfies the necessary bounds
\cite{mo:bio2}
\begin{equation}
\frac{1}{2}\leq\rho_1(\omega,\alpha)\leq 1
\end{equation}
in the $\omega-$interval

\begin{equation}
\arccos\left(\frac{1}{3}\right)\leq\omega\leq
2\pi-\arccos\left(\frac{1}{3}\right)
\end{equation}

If we wish to restrict the $\rho$-values by using the Du-Hwang's
greatest lower bound $(\rho=\sqrt{3}/3)$ instead of Moore's
$(\rho=1/2)$, we have to substitute (6.2) by

\begin{equation}
\arccos\left(\frac{1}{4}\right)\leq\omega\leq
2\pi-\arccos\left(\frac{1}{4}\right)
\end{equation}

We need a constraint to be motivated by the natural organization
of a macromolecular structure in our modelling. Some interesting
results have been obtained \cite{mo:bio3} by introducing a
function of the form
\begin{equation}
H(\omega,\alpha)=(1+\lambda)\rho_1(\omega,\alpha)-\lambda\phi(\omega,\alpha)
\end{equation}
where $\lambda$ is a Lagrange multiplier and $\phi(\omega,\alpha)$
stands for a recently proposed function for chirality measure
\cite{mo:bio3} given by

\begin{equation}
\phi(\omega,\alpha)=\frac{1}{6}\alpha \omega
\sin(\omega)\left(\frac{A_1+1}{A_1}\right)(\alpha^2\omega^2-A_1(A_1+1)).
\end{equation}

The work with (6.4) has produced a new upper bound for
$\rho_1(\omega,\alpha)$ which is still lower than the
unconstrained minimum of $\rho_1(\omega,\alpha_{\mathrm{R}})$ as
compared to the Global minimum value of $\rho(\omega,\alpha)$.

It would be easy to announce a proof for the conjecture of
\cite{du:disp} or at least some very good arguments for a proof
from the work of the last sections. However, in spite of the
evidences which were reported here, we think that the approach
adopted has given nice advances in our program of modelling the
structure of biomacromolecules with Steiner Points and Steiner
Trees. The possibility of analyzing the chirality effects on these
structures is one of these advances.


\begin{thebibliography}{99}

\bibitem{sm:tree}
Smith, J. M.: Steiner Minimal Trees in $E^3$: Theory, Algorithms
and Applications. Handbook of Combinatorial Optimization
{\bfseries 2}, Kluwer Acad. Publ. (1998) 397--470.

\bibitem{mo:bio1} Mondaini, R. P.: The Minimal Surface Structure of Biomolecules.
Proceedings of the First Brazilian Symposium on Mathematical and
Computational Biology -- ed. E-papers LTDA, Rio de Janeiro (2001)
1--11.

\bibitem{mo:bio2}
Mondaini, R. P.: The Disproof of a Conjecture on the Steiner Ratio
in $E^{3}$ and its Consequences for a Full Geometric Description
of Macromolecular Chirality. Proceedings of the Second Brazilian
Symposium on Mathematical and Computational Biology -- ed.
E-papers LTDA, Rio de Janeiro (2002) 101--177.

\bibitem{mo:bio3}
Mondaini, R. P.: Proposal for Chirality Measure as the Constraint
of a Constrained Optimization Problem. Proceedings of the Third
Brazilian Symposium on Mathematical and Computational Biology --
ed. E-papers LTDA, Rio de Janeiro (2004) 65--74.

\bibitem{mo:bio4} Mondaini, R. P.: The Geometry of Macromolecular
Structure: Points and Steiner Trees. Proceedings of the Fourth
Brazilian Symposium on Mathematical and Computational Biology --
ed. E-papers LTDA, Rio de Janeiro (2005).

\bibitem{mo:kluw}
Mondaini, R. P.: The Steiner Ratio and the Homochirality of
Biomacromolecular Structures. Frontiers in Global Optimization:
Nonconvex Optimization and its Applications Series {\bfseries 74},
Kluwer Acad. Publ. (2003) 373--390.

\bibitem{si:algo}
Smith, W. D.: How to find Steiner Minimal Trees in Euclidean
$d-$Space. Algorithmica {\bfseries 7} (1992) 137--177.

\bibitem{sm:ratio}
Smith, W. D., MacGregor Smith, J.: The Steiner Ratio in 3D Space.
Journ. Comb. Theory {\bfseries A69} (1995) 301--332.

\bibitem{mo:gmb}
Mondaini, R. P.: The Euclidean Steiner Ratio and the Measure of
Chirality of Biomacromolecules. Genetics and Molecular Biology,
vol.27, {\bfseries 4} (2004) 658--664.

\bibitem{si:tre2}
Smith, J. M., Toppur, B.: Euclidean Steiner Minimal Trees, Minimal
Energy Configurations, and the Embedding Problem of Weighted
Graphs in $E^3$. Discret. Appl. Math. {\bfseries 71} (1996)
187--215.

\bibitem{mo:tema}
Mondaini, R. P., Oliveira, N. V.: The State of Art on the Steiner
Ratio Value in $\bbbr^{3}$. Tend\^encias em Matem\'atica Aplicada
e Computacional, vol.5, {\bfseries 2} (2004).

\bibitem{du:disp}
Du, D. Z., Smith, W. D.: Disproofs of the Generalized
Gilbert-Pollak Conjecture on the Steiner Ratio in Three or more
Dimensions. Journ. Comb. Theory {\bfseries A74} (1996) 115--130.

\bibitem{mk:prog}
Bazaraa, M. S., Sherali, H. D., Shetty, C. M.: Nonlinear
Programming. Wiley (1993) pp.125.
\end{thebibliography}
\end{document}